\documentclass[twocolumn,showpacs,preprintnumbers,amsmath,amssymb,a4paper,floatfix]{revtex4}

\usepackage{graphicx}
\usepackage{dcolumn}
\usepackage{bm}

 \newcommand{\T}[1]{\mathrm{#1}}
 
 \newcommand{\C}{^{\circ}\mathrm C}
 \newcommand{\AR}{\rightarrow}
 \newcommand{\DS}{\displaystyle}
 \newcommand{\D}{\Delta}

\begin{document}

\title{Self-consistent Monte-Carlo simulation
       of \\ the positive column in Argon }

 \author{V.Zhakhovskii, and K.Nishihara}
 \affiliation{ILE, Osaka University.}
 \email{basil@ile.osaka-u.ac.jp}

\date{\today}

\pacs{52.80.Hc, 34.80}

\begin{abstract}
A high accurate self-consistent Monte-Carlo method including charged and neutral particle motion and resonance
photon transport is developed and applied to simulation of positive column discharge in Argon. The distribution
of power loss across the tube is investigated. It is shown that negative slope of current-voltage gradient
curve can not be explained by only one factor. Diffusion of metastable atoms to the tube wall in the case of
low current ($0.2-3$mA) responds for the most part of negative slope. For the high currents the variation of
radiation power loss makes a main contribution to the amplitude of negative slope.
\end{abstract}

\maketitle

\section{Introduction}

The positive column of gas discharge has been experimentally studied for a long time but still there are some
lack of the simulation techniques which are able to reproduce discharge properties in closed details.

Modeling of the positive column (PC) is based mostly on fluid approaches, which are inapplicable for sheath
region as well for low pressures and small currents because they do not reflect the physical picture of
discharge (due to assumption of Maxwellian electron distribution function and simplified treatment of photon
transfer) and results in inadequate simulation.

As it was well established, electron component of discharge generally reaches a state far from the
thermodynamic equilibrium. Hence the temporal and spatial distributions of anisothermic plasmas can be
described only on appropriate microphysical basis. There are two quite different approaches. One way consists
of the formulation of a kinetic equation and its approximate solution. Until recently, most studies of the
electron behavior in positive columns based on solution of the Boltzmann equation \cite{Winkler1}. The other
approach uses the technique of the particle-in-cell simulation coupled with Monte-Carlo (MC) event selection.
The last method has no disadvantages of the kinetic Boltzmann method, and it is limited only the available
computing power \cite{Lawler, Lee2001}.

Electron and ion motions in low temperature plasmas may be studied by MC approach in which the path of a single
particle is traced by computer simulation. The particle is considered to move classically during period of free
flight under influence of a self-consistent electric field. This field is determined by electron and ion
density according to Poison's law. The free flight is interrupted with scattering by neutral atoms, ionization,
excitation, wall loss and other kind of collisions which must be considered in the model of discharge. The free
flight length and outcome of these events are selected randomly from known probability distributions. Transport
properties of the discharge under investigation are found from the time average of the behavior of the
individual particle. When an collision occurs one of the possible collision mechanisms must be chosen. The
probability of each mechanism is proportional to its frequency. Then the particle velocity is modified
according to the physical nature of the mechanism chosen. The simulation of a number of particles may be
carried out simultaneously and the time dependence of their behavior can be derived from an ensemble average.

Resonance photon transfer is also very important for a quantitative simulation of gas discharge tube as well
the charged particle motion. A new simple Monte-Carlo treatment of radiation imprisonment based on
Holstein-Biberman theory \cite{Holstein} is developed and embedded into the model.

We present a high accurate self-consistent Monte-Carlo approach to simulation of the positive column of Argon
discharge in this short report.

\section{ Particle transport processes}

At present model the first 14 excited Ar levels are involved into consideration. The 15th level of Argon
$\T{Ar}_{15}(3d_a)$ is a combined level that represents great number of the higher excited atomic levels as one
state. We assume that Ar may be ionized and excited or de-excited by electron-atom collisions according to
general formula which can be written as:
\begin{equation}
      e + \T{A}_i \AR \sigma(i,j,u) \AR e + \T{A}_j,          \label{eq:PT02}
\end{equation}
where A is atomic symbol, index $i$ denotes initial state (before collision), $j$ corresponds to the final
state, and $\sigma(i,j,u)$ is a cross section for given pair $(i,j)$ and electron kinetic energy is $u$. In the
case of $i<j$ Eq.(\ref{eq:PT02}) represents excitations and ionization of atom, elastic scattering in the case
of $i=j$, and super-elastic electron-atom collisions for $i>j$.

The electron excitation and ionization cross sections are taken from Refs.\cite{Hayashi1, Bartschat}. After
ionization of atom a new electron appears. We assume that both the ejected and scattered electrons are
redistributed isotropically, and both electrons randomly share available energy which is given by subtracting
the ionization energy for a given energy level from the kinetic energy of the incident electron. At very low
pressure of Argon the excitation of inner electron shells of atom and multiple ionization becomes more probable
due to the higher electron temperature. The present model contains very simplified calculation of double
ionization events and may obtain only qualitative description of discharges at electron temperature $T_e
> 10$eV.

The cross section of the super-elastic collisions $\sigma'(j,i;u')$  (reverse reaction to Eq.(\ref{eq:PT02}))
is calculated according to the principle of detailed balance \cite{Lieberman}:
\begin{equation}
    u g_i \sigma(i,j;u) = u' g_j \sigma'(j,i;u')                   \label{eq:PT03}
\end{equation}
where $j>i$ and $ u' = u - (E_j-E_i) > 0$ is initial electron kinetic energy, $g$ is a degeneracy of the energy
level. Therefore the inverse cross section
\begin{equation}
 \sigma'(j,i;u') = \frac{g_i}{g_j} \left(1+\frac{E_j-E_i}{u'}\right) \sigma(i,j; u'+(E_j-E_i)).  \label{eq:PT04}
\end{equation}

 \begin{figure}
 \centering
 \includegraphics [width=\columnwidth] {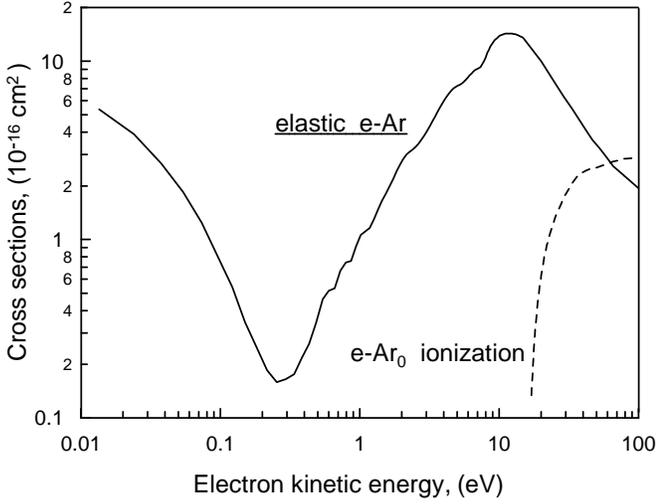}
 \caption{ \label{fig:els_e-Ar} Solid line correspond to momentum transfer cross section for electrons in Argon
                                \cite{Yousfi}.  Note the Ramsauer minimum below 1 eV. Dash line denotes electron
                                impact ionization cross section from ground state of Ar.}
 \end{figure}
In the MC model we consider elastic electron-Ar scattering by using the published momentum transfer cross
section \cite{Yousfi} (see Fig.\ref{fig:els_e-Ar}). On average an elastic recoil energy loss of $2m_e/M
=2.73\times 10^{-5}$ of the incident kinetic energy are used in all simulations. Here $m_e$ is electron mass
and $M$ is Argon atom mass. After elastic collision an electron is redistributed isotropically in the
center-of-mass (CM) system.

It is often supposed in positive column models that all ions and electrons reaching the tube wall recombine one
another. This assumption is also used in our model. Recombination of ions at wall produces the neutral atoms in
ground state. The program redistributes uniformly these atoms within tube domain to maintain balance of the
atom number density.

\begin{figure}
\centering
\includegraphics [width=\columnwidth] {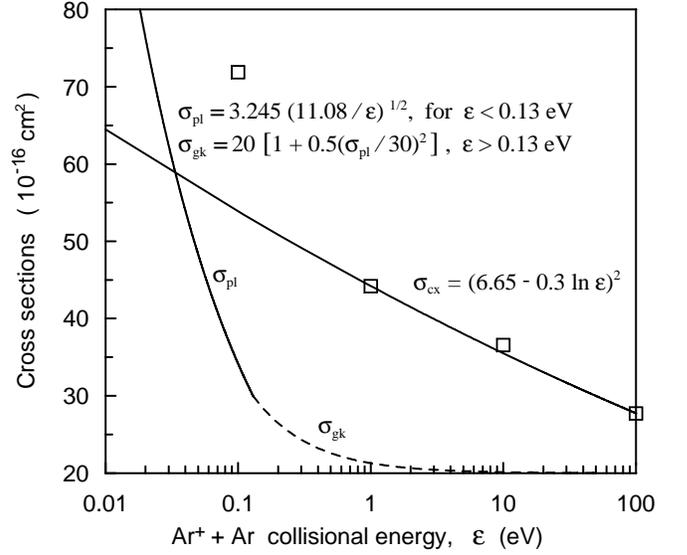}
\caption{\label{fig:els_i-Ar} Momentum transfer cross section and resonant charge transfer
                              cross section for $\T{Ar^+}$ in parent gas.
                              Lines are fitted functions, symbols denote experimental data \cite{Sprav}.}
\end{figure}

In contrast to elastic electron-atom collisions where energy exchange is limited by factor $2m_e/M \ll 1$,
energy exchange in electron-electron collision has the same order as energy of colliding electrons and moreover
a huge electron-electron momentum transfer cross section (Coulomb cross section). For low axial electric field
and not too small current, the electron-electron collisions result mainly in energy transfer from cold bulk
electrons to hot tail electrons which are responsible for particle (ionization) balance in discharge. So the
correct treatment of electron-electron collisions is one of the key elements of the discharge simulation.

To describe electron-electron collisions by MC simulation we are using classical Coulomb cross section
\cite{Weng}:
\begin{equation}
   \sigma(v) = 4\pi b_0^2 \left( 1+ \ln(\lambda_D /b_0)^2 \right)^{1/2}, \label{eq:PT05}
\end{equation}
where $\lambda_D=\left(kT_e/4 \pi e^2 n_e \right)^{1/2} $ is Debye length  and $b_0=e^2/\mu v^2=2e^2/mv^2$ is
Coulomb radius, $\mu$ is the reduced mass of colliding electrons, $v$ is their relative speed, and $n_e, T_e$
are the local electron density and temperature. The energy exchange during particle-particle (here
electron-electron) scattering depends on the velocity vectors of a projectile 1st-particle and target
2nd-particle according to equations:
\begin{equation}
 \mathbf{v}_1^\prime=
 (m_1\mathbf{v}_1 + m_2\mathbf{v}_2 + m_2 v\mathbf{n})/(m_1+m_2)    \label{eq:PT06a}
\end{equation}
\begin{equation}
 \mathbf{v}_2^\prime=
 (m_1\mathbf{v}_1 + m_2\mathbf{v}_2 - m_1 v\mathbf{n})/(m_1+m_2)    \label{eq:PT06b}
\end{equation}
where $\mathbf{v}_1^\prime$ and $\mathbf{v}_2^\prime$ are the velocities of particles in the laboratory system
after collision. The unknown unit vector $\mathbf{n}$ is assumed to distribute isotropically.

Calculation of electron-electron collisional frequency takes much larger computational power than
electron-heavy particles collisional frequency because of target electron distribution is unknown a priori. Let
us consider electron collision frequency $\nu_i$  for some $i$-projectile electron moving with given velocity
$\mathbf{v}_i$ in a spatial domain $\mathbb{V}$ given by a computational mesh:
\begin{equation}
  \nu_i(\mathbf{v}_i)= n_e\int |\mathbf{v}_i-\mathbf{v}_j| \;
                     \sigma(|\mathbf{v}_i-\mathbf{v}_j|)
                      f_V(\mathbf{v}_j) d \mathbf{v}_j                  \label{eq:PT07}
\end{equation}
where subscript $j$ is the number of target electrons, $f_V(\mathbf{v})$ is local electron distribution
function of the domain $\mathbb{V}$ and $n_e=N/\mathbb{V}$ is local electron density of the number of electrons
$N$ within $\mathbb{V}$.  For relative velocity $v_{ij}=|\mathbf{v}_i-\mathbf{v}_j|$:
\begin{equation}
  \nu_i(\mathbf{v}_i)= n_e\int v_{ij} \; \sigma(v_{ij}) f_V(\mathbf{v}_j) d \mathbf{v}_j=
  n_e \left\langle v_{ij} \; \sigma(v_{ij}) \right\rangle_i                              \label{eq:PT08}
\end{equation}
Using velocity data of target particles Eq.(\ref{eq:PT08}) can be written as follows
\begin{equation}
 \nu_i(\mathbf{v}_i)=\frac{N}{\mathbb{V}} \frac{\sum_j^N     v_{ij}\;\sigma(v_{ij})}{N}  \cong
                    \frac{N}{\mathbb{V}} \frac{\sum_j^{N_j} v_{ij}\;\sigma(v_{ij})}{N_j} \label{eq:PT09}
\end{equation}
where $N_j$ is number of accounted targets. To reduce computational efforts one may use $N_j\ll N$ as shown in
the last part of Eq.(\ref{eq:PT09}).

We assume in our model that the local collisional frequency $\nu(\mathbf{v}_i)$ in Eq.(\ref{eq:PT09}) depends
only on electron kinetic energy $u=m\mathbf{v}^2/2$ instead of vector velocity $\mathbf{v}$.  Then the
collisional frequency $\nu(u_k)$ for electrons having kinetic energy in the $k$ energy bin $(u_k,u_k+\Delta
u_k)$ becomes
\begin{equation}
\nu(u_k)= \frac{\sum_i^N \nu_i(\mathbf{v}_i) \delta(u_i,u_k)}{\sum_i^N  \delta(u_i,u_k)} =
\frac{N}{\mathbb{V}}\frac{\sum_{i,j}^{N,N_j}v_{ij}\;\sigma(v_{ij})\delta(u_i,u_k)}{N_k N_j} \label{eq:PT10}
\end{equation}
here $\delta(u_i,u_k)=1$ if $u_i\in(u_k,u_k+\Delta u_k)$, and $\delta(u_i,u_k)=0$ if $u_i\notin(u_k,u_k+\Delta
u_k)$, $N_k=\sum_i^N \delta(u_i,u_k)$ is number of projectile electrons in $k$-bin.

The electron-ion and ion-ion elastic collision processes are also included to the model in the same manner as
for electron-electron collision. However according to our simulation these processes do not play any
significant role in the PC.

It should be note that for arbitrary projectile-target pair the relative velocity may be a very small value and
the corresponding Coulomb radius $b$ in Eq.(\ref{eq:PT05}) becomes larger than Debye length and/or spatial mesh
size. In these cases the electron-electron collision event is assumed as a null collision because the long
range electron interactions are already included in solution of Poisson's equation.

%
Ion transport in the ion-parent-atom case is mostly guided by resonant charge transfer because of its cross
section is large at low collision energies. We use a simple fitted function for the charge transfer cross
section \cite{Jonsen,Chanin} given by
\begin{equation}
 \sigma_{cx} = (A - B \ln \varepsilon)^2                       \label{eq:PT11}
\end{equation}
Here $\varepsilon$[eV] is a kinetic energy of a projectile ion in CM system, $A$ and $B$ are the fitting
parameters: $A_{Ar}= 6.65$[\AA] , $B_{Ar}= 0.3$[\AA], see Fig.\ref{fig:els_i-Ar}. Due to the large cross
section the ions and atoms are practically undeflected in the center-of-mass system. Therefore, after the
resonant charge transition occurs a new ion will accept velocity of the neutral atom which are randomly
generated according normal distribution function for a given gas temperature which is assumed to be constant
across the tube.

Polarization scattering of ion by a neutral atom can be described by momentum transfer cross section
\cite{Raizer} (without $2\sqrt{2}$ term):
\begin{equation}
   \sigma_{pl}= \pi a_0^2 \sqrt{(\alpha/a_0^3)I_H/\varepsilon} ,   \label{eq:PT12}
\end{equation}
where $a_0=\hbar^2/me^2 =0.5292\cdot 10^{-8}$cm is Bohr radius, $I_H=e^2/2a_0=13.6$eV is ionization potential
of Hydrogen atom, and $\varepsilon=M' v^2/2$ is kinetic energy of relative motion of ion and atom (that is in
the CM system). The polarizability of Argon atom is equal to $\alpha_{Ar}/a_0^3=11.08$. The minimal value of
$\sigma_{pl}$ is limited by the corresponding gas-kinetic (hard sphere) cross section $\sigma_{gk}$ as shown in
Fig.\ref{fig:els_i-Ar}. We assume that the ion is redistributed isotropically in CM system according to
Eqs.(\ref{eq:PT06a},\ref{eq:PT06b}) after an elastic collision with a random neutral atom at a given gas
temperature $T_g=20 \C$.

Because the correct calculation of ion motion is crucial point for the gas discharge modeling, we carry out the
validity check of these cross sections by comparing the simulated ion drift velocity with experimental one.
Reduced mobility of $\T{Ar^+}$ in argon gas is $\mu_0(\T{Ar^+,Ar})=1.6 \;\T{cm^2 V^{-1}s^{-1}}$ \cite{Jonsen},
hence the experimental drift velocity for gas pressure 0.261 Torr and electric field 2 V/cm is 93.18 m/s which
is in good agreement with simulated $\T{Ar^+}$ drift velocity 92.3 m/s.

%
We take into account excited atom-atom collision ionization process as follow $ \T{Ar}_i^*+\T{Ar}_j^* \AR
\T{Ar}_0+\T{Ar}^+ + e $, where $\T{Ar}^*$ is any one of a number of excited Ar levels. Cross sections of these
processes found in the literature vary from $10^{-20}\T{m}^2$ to $100\times 10^{-20}\T{m}^2$ \cite{Vriens78}.
We assume the recommended value $24\times 10^{-20}\T{m}^2$ \cite{Zissis}. The new ionized electron gains
kinetic energy $E_i+E_j-I_{Ar}>7.86$[eV] and the new Ar ion accepts random velocity from a neutral Ar atom.

The simple treatment of diffusion based on equation of continuity is included to the model. The flux of atoms is
estimated on basis of Fick's law given by
\begin{equation}
   j_d = -D \nabla n(r), \;\; D=D_0(P_0/P)(T/T_0)^{\alpha} \;,      \label{eq:PT13}
\end{equation}
here $D_0=0.073 \;\T{cm^2/s}$ is diffusion coefficient of the excited Ar atoms in Argon gas at the normal
conditions, and $\alpha=1.92$ \cite{Sprav}.

In case of a large gradient of density the diffusion flux $j_d$ may be larger than kinetic limit of the flux of
particles with averaged velocity $\langle V_a\rangle $ given by
\begin{equation}
   j_k = n\langle V_a\rangle /4                                     \label{eq:PT14}
\end{equation}
To eliminate the overestimation of diffusion processes in Eq.(\ref{eq:PT13}) we assume the total flux as the
following formula:
\begin{equation}
   j = j_k j_d/(j_k+j_d)                                           \label{eq:PT15}
\end{equation}

The charged particle is considered to move classically during period between collisions (free flight) under
influence of an self-consistent electric field within discharge tube. This field is determined by electron and
ion density according to Gauss's law:
\begin{equation}
 \oint {\mathbf E} d {\mathbf S} = 4 \pi Q = 4 \pi \int (n_i-n_e)dV   \label{eq:PT16}
\end{equation}
Because we suppose particle densities are symmetric about tube axis, the electric field ${\mathbf E}$ in the
cylindrical coordinate system has only the radial component $E_r(r)$. It provides us the simple formula
\begin{equation}
      E_r(r) = 2Q(r)/r                                              \label{eq:PT17}
\end{equation}
where $Q(r)$ is the total charge within radius $r$ per unit length of discharge tube.

%
              \section{ Resonance radiative transport }

\begin{figure}
\centering
\includegraphics [width=\columnwidth] {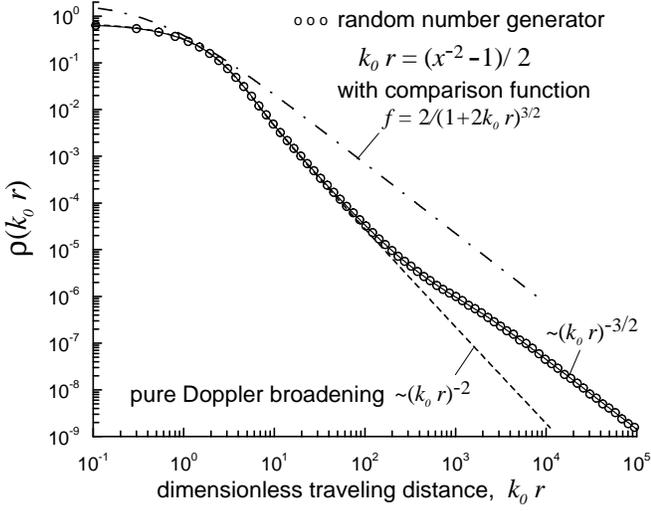}
\caption{\label{fig:Voigt_0186} The transmission probability density at Voigt parameter a = 0.0186.
                                Solid line denotes theoretical calculation by Eq.(\ref{eq:RT05}). }
\end{figure}

Due to the high absorption of the resonance photons by atoms in the ground state the correct treatment of
radiative transfer in discharge condition is very important for simulation model. We apply the
Biberman-Holstein theory \cite{Holstein} of resonance radiative transport for MC simulation of traveling
photons.

Let us denote $\rho(r)=k\exp(-kr)$ the probability density of a photon captured within $\left[r,r+dr \right]$
and without absorption along distance $\left[0,r \right]$. Then the probability of the absorption anywhere
within $\left[0,R \right]$ is $\int_0^R \rho(r)dr = 1-\exp(-kR)$  and  probability of the traversing a distance
$R$ is $T(R)=1-\int_0^R \rho(r)dr = \exp(-kR)$. Due to line broadening (natural, pressure and Doppler
broadening) the absorption coefficient $k$ depends on a photon frequency $\omega$.

For the pure Doppler broadening case since the velocity distribution of the atoms is a Maxwellian with given gas
temperature, one can obtain the absorption coefficient as
\begin{equation}
 k(\omega)=k_0\exp[-(\omega/\omega_0-1)^2 c^2 m/2T_g],           \label{eq:RT01}
\end{equation}
where $k_0$ is absorption coefficient in the maximum of the line profile, $c$ is speed of light, $m$ is an
atomic mass, and $T_g$ is a gas temperature in energetic units.

It can be shown that the averaged probability density of a photon capture after traveling distance $r$ is
\begin{equation}
 \rho(r)=\frac{k_0}{\sqrt{\pi}}
 \int\exp\left[-2x^2 - k_0 r e^{-x^2}\right]dx                   \label{eq:RT02}
\end{equation}
here $x=\frac{\DS\omega-\omega_0}{\DS\omega_0} \: c/(2T_g/m)^{1/2}$, and $k_0$ is given by
\begin{equation}
 k_0= \frac{\lambda_0^3}{8 \pi^{3/2} } \frac{g_2}{g_1}
      \frac{ n_A }{ \tau(2T_g/m)^{1/2} }                            \label{eq:RT03}
\end{equation}
Here $\tau$ is mean radiative lifetime of an excited level , $\lambda$ is wavelength, and $n_A$ is number
density of atoms. We assume that the natural argon gas consists of only one isotope with atom mass 39.948 a.u.
For given set of 15 atomic levels we apply the natural mean life times as they are listed in NIST Atomic
Transition Probability Tables \cite{CRC} (see page 10-88).

In a general case the emission (or absorption) line shape is determined not only by the Doppler broadening but
natural and pressure (Lorentz) broadening as well. The combination of these effects may be given by Voigt
profile $h(a,x)$ \cite{Payne}:
\begin{equation}
  h(a,x)= \frac{a}{\pi} \int\limits_{-\infty}^{\infty}
                           \frac{\exp(-y^2)dy}{a^2+(x-y)^2},   \label{eq:RT04}
\end{equation}
and then the emission profile $\varphi(x)$ and the absorption coefficient $k(x)$ are written as
\begin{equation}
  \varphi(x)= h(a,x)/\sqrt{\pi}, \qquad  k(x)= k_0 h(a,x).
\end{equation}
Here $x$ was defined above and $a$ is the Voigt parameter (ratio of the Lorentz HWHM to Doppler width at 1/e
maximum) defined below
\begin{equation}
 a=\frac{\lambda_0}{4 \pi \tau (2T/m)^{1/2} }
                   \left( 1 + 2n_A\tau\langle v_{AA}\sigma_A\rangle \right),
\end{equation}
where $\sigma_A$ is a resonance broadening cross section which is $\sigma_A=643$\AA$\!^2$ for 106.67 nm Ar
line, and $\sigma_A=2339$\AA$\!^2$ for 104.82 nm Ar line in the pure Argon discharge. In the program we use the
alternate formula for resonance broadening \cite{Payne} given by
\begin{equation}
\tau \langle v_{AA}\sigma_A\rangle = 0.4506 \: \frac{g_2}{g_1} \frac{\lambda_0^3}{6 \pi^2}
\end{equation}
The Voigt parameters for Argon discharge at pressure 0.261 Torr are $a=0.0172$ for $\lambda_0=104.82$nm, and
$a=0.00468$ for $\lambda_0=106.67$nm.

The function $h(a,x)$ from Eq.(\ref{eq:RT04}) has two asymptotic limits for $a=0$ and $a\AR\infty$. The pure
Doppler broadening is realized as $h(a,x)\AR \exp(-x^2)$ when $a\AR 0$. In the case of $a\AR\infty$ the Voigt
profile tends to Lorentz profile as $h(a,x)\AR a/\sqrt{\pi}(a^2+x^2))$.

Similarly to Eq.(\ref{eq:RT02}) the transmission probability density in the most general form is given by
\begin{equation}
 \rho(r)= \int \varphi(x) k(x) e^{-k(x)r} dx
\end{equation}
and can be rewritten for the Voigt profile as follow
\begin{equation}
 \rho(k_0 r)= \frac{k_0}{\sqrt{\pi}} \int h^2(a,x)\exp{[(-k_0 h(a,x)r ]}dx  \label{eq:RT05}
\end{equation}

To generate a random number according to the probability density $\rho(k_0 r)$ from the previous Equation we
develop a new simple random generator
\begin{equation}
     k_0 r = \left(X^{-2} - 1 \right)/2                            \label{eq:RT06}
\end{equation}
with comparison function
\begin{equation}
    f(k_0 r) = 2/(1+2k_0r)^{3/2}                                   \label{eq:RT07}
\end{equation}
At first step the program generates a trial random number $k_0 r$ according to Eq.(\ref{eq:RT06}) by using a
random number $X$ uniformly distributed within $[0,1]$. At second step the uniformly distributed within
$[0,f(k_0 r)]$ random number $Y$ is picked over. If $Y>\rho(k_0 r)$ then the program rejects the trial number
$k_0 r$ and return to the first step. Otherwise the program accept the trial number. The efficiency of this
generator is equal 50\% because of $\int f(k_0 r) d(k_0 r)=2$. Figure \ref{fig:Voigt_0186} indicates that the
probability density generated by random generator Eq.(\ref{eq:RT07}) is in a excellent agreement with
theoretical function from Eq.(\ref{eq:RT05}). The transmission probability density $\rho(k_0 r)$ is tabulated
and stored at first start of the program.

%
\section{ Particle balance theory}

To maintain the charge balance in a steady PC an electron may produce many excited atoms and only one
electron-ion pair during the electron mean life time. The exited atoms may disappear by radiative decay, in
electron inelastic or super-elastic collisions, in electron impact and atom-atom ionization processes, and due
to diffusion to the tube wall. In our model the ions can only recombine at the tube surface. It is clear that
the total balance of each sort of particles must be equal zero in the steady positive column.

As example of simplified theoretical model of PC let us consider the electron balance equation which can be
written as sum of the volume rates:
\begin{equation}
  \sum_i \nu_{ie} n_e  + \sum_{ij} k_{ij} n_i n_j - \nu_{ew} n_e = 0  \label{eq:Pb01}
\end{equation}
where first term corresponds to direct ionization from i-level atoms by electron impact with frequency
$\nu_{ie}=k_{ie}n_i$ in unit volume, second term denotes atom-atom ionization in collisions between excited
i-level atom and j-level atom, and third term is electron loss rate on the tube wall.

In the simplest theory of DC positive column (Schottky, 1924) the electron-wall loss frequency $\nu_{ew}$ is
assumed to be independent on electron density $n_e$ and equal to $\nu_{ew}= D_e (2.405/R)^2$, where $D_e=\mu_i
T_e$ is the ambipolar diffusion coefficient and $T_e$ weakly depends upon the electron density. In the simplest
model of positive column where the ionization from ground state is only taken into consideration the ionization
frequency $\nu_{0e}$ does not depend on electron density and ionization rate $\nu_{0e}n_e$ is linear term with
respect to $n_e$. The electron balance equation Eq.(\ref{eq:Pb01}) becomes simply $\nu_{0e} = \nu_{ew}$,
therefore electron balance does not depend on electric current which is proportional to $n_e$. Ionization
frequency $\nu_{0e}$ is a function of the axial electric field $E_z$. Hence $E_z$ is independent of electric
current and only determined by electron wall-loss frequency.

We may extend Schottky theoretical model by including ionization from one sort of the excited resonance atoms
only as it have been done in \cite{Sommerrer}:
\begin{equation}
  k_{0e}n_0 n_e + k_{xe}n_x n_e - \nu_{ew} n_e = 0              \label{eq:Pb02a}
\end{equation}
\begin{equation}
  k_{0x}n_0 n_e - k_{xe}n_x n_e - \nu_{x0} n_x = 0              \label{eq:Pb02b}
\end{equation}
where $\nu_{x0}=1/\tau_x$ is radiative decay frequency of x-level atom to ground state 0, $n_x$ is number
density of excited atoms. By solving Eq.(\ref{eq:Pb02b}) for $n_x$ and rewrite Eq.(\ref{eq:Pb02b}) to obtain
balance equation
\begin{equation}
 k_{0e}n_0n_e\left(1 + \frac{n_e}{ k_{xe}n_e + \nu_{x0}}\right) -\nu_{ew} n_e = 0  \label{eq:Pb03}
\end{equation}
It is clear that the ionization term in Eq.(\ref{eq:Pb03}) is superlinear in $n_e$, except the case $ k_{xe}n_e
\gg \nu_{x0}$ (high current). For this model we may conclude that the ionization from resonance states results
in negative slope of V-I characteristic curve \cite{Sommerrer}. It is interesting to note that for high current
case and/or for metastable atom with $\nu_{x0}=0$ the balance equation (\ref{eq:Pb03}) gives a flat V-I curve.

Let us consider in contrast to \cite{Sommerrer} the same model Eqs.(\ref{eq:Pb02a},\ref{eq:Pb02b}) but now for
metastable x-level atoms with $\nu_{x0}=D_x (2.405/R)^2$ as frequency of diffusion loss on the tube wall:
\begin{equation}
  k_{0e}n_0 n_e + k_{xe}n_x n_e - \nu_{ew} n_e = 0  \label{eq:Pb04a}
\end{equation}
\begin{equation}
  k_{0x}n_0 n_e - k_{xe}n_x n_e - \nu_{x0} n_x = 0  \label{eq:Pb04b}
\end{equation}
and we get the similar solution given by
\begin{equation}
 k_{0e}n_0n_e\left(1 + \frac{n_e}{ k_{xe}n_e + \nu_{x0}}\right) -\nu_{x0} n_e = 0  \label{eq:Pb05}
\end{equation}
For this model in the enough low current case ($k_{xe}n_e < \nu_{x0}$) the ionization from metastable states
causes a negative V-I slope. Because the number of metastable atoms is much higher than the number of resonance
atoms in low current region Eq.(\ref{eq:Pb05}) conforms to the discharge much better than
Eqs.(\ref{eq:Pb02a},\ref{eq:Pb02b}). Hence we may conclude the diffusion of metastable atoms to the tube wall
is an essential requirement for a negative V-I characteristic of the low current positive column.

The atom-atom ionization should be included in any realistic model. Let us insert this atom-atom ionization
term $k_{xx}n_x^2$ into our model Eqs.(\ref{eq:Pb04a},\ref{eq:Pb04b}):
\begin{equation}
  k_{0e}n_0 n_e + k_{xe}n_x n_e +  k_{xx}n_x^2 - \nu_{ew} n_e = 0      \label{eq:Pb06a}
\end{equation}
\begin{equation}
  k_{0x}n_0 n_e - k_{xe}n_x n_e - 2k_{xx}n_x^2 - \nu_{x0} n_x = 0      \label{eq:Pb06b}
\end{equation}
By combination of two equations it is easy to get the final electron balance formula:
\begin{equation}
 k_{0e}n_0 n_e  + k_{xe} w n_e^2 + k_{xx} w^2 n_e^2 -\nu_{ew} n_e = 0  \label{eq:Pb07}
\end{equation}
where
\begin{equation}
 w = \frac{ 2\nu_{ew}-(2k_{0e}+k_{0x})n_0 } { k_{xe}n_e-\nu_{x0}  }
\end{equation}

\begin{figure}
\centering
\includegraphics [width=\columnwidth] {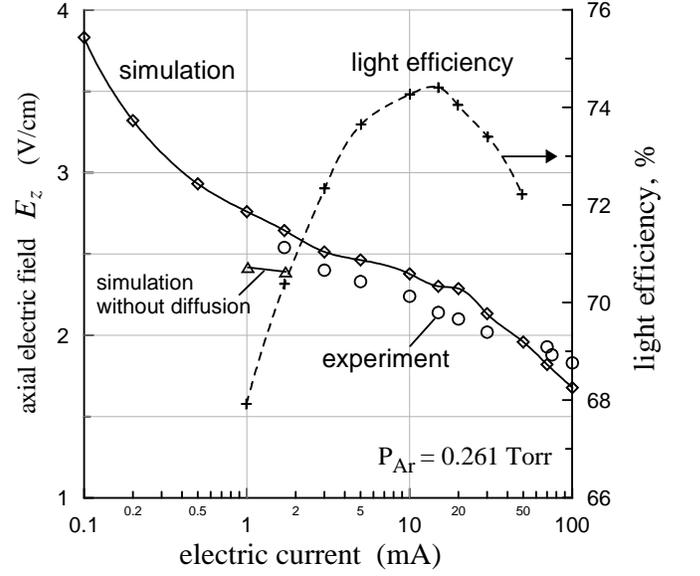}
\caption{\label{fig:EI_0.261Tor} Voltage-current characteristic of Ar positive column at pressure 0.261 Torr,
             and the efficiency of light emission to the total power dissipated in PC. Squares denote simulation
             data and the circles show the experimental points. Triangles corresponds to simulation without
             diffusion of excited atoms to the tube wall.}
\end{figure}

\begin{figure}
\centering
\includegraphics [width=\columnwidth] {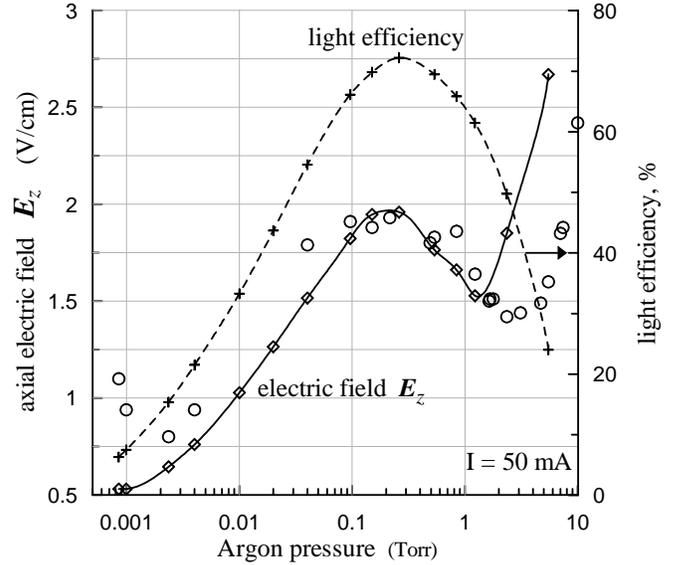}
\caption{\label{fig:EP_50mA} The voltage-pressure characteristic curve of Ar PC at current $I=50$ mA.
                             See notations at Fig.\ref{fig:EI_0.261Tor}.}
\end{figure}

Again for the enough low current case ($k_{xe}n_e < \nu_{x0}$) Eq.(\ref{eq:Pb07}) shows two superlinear
ionization terms, the first is atom-atom ionization and the second is electron impact ionization from
metastable atoms. These superlinear terms result in negative slope of V-I curve. But with increasing of current
the atom-atom ionization term $k_{xx} w^2 n_e^2$ becomes a sublinear term in $n_e$ (almost independent of
$n_e$), but the impact ionization from metastable atoms tends to linear term in $n_e$. It may cause a positive
slope of V-I curve if $k_{xe} < k_{xx}w.$ For further increasing of current we have to take in consideration
the electron impact ionization from resonance states (see Eq.(\ref{eq:Pb04a},\ref{eq:Pb04b})) and transitions
between metastable and resonance states. It adds to Eq.(\ref{eq:Pb07}) a new superlinear ionization term which
have to turn V-I curve from the growth to the drop of voltage. Therefore the unknown whole electron balance
equations have to contain the ionization rates consisting from the sublinear, linear, and superlinear terms as
we show above. The interplay between them results in a slope of V-I curve. We may distinguish at least three
electric current region in accordance with slope sign of the V-I curve. Low current region (ionization from
metastable atom $k_{xe}n_e < \nu_{x0}$) corresponds to the negative slope, the middle current region is for the
positive slope (due to independence of atom-atom ionization rate on $n_e$), and the high current region
corresponds again to the negative slope ( due to superlinear ionization from excited atoms $k_{xe}> k_{xx} w$).

We have to note that this theoretical model is not applicable for very low current PC where the electron-wall
loss frequency $\nu_{ew}$ depends on electron density $n_e$, and the electron temperature significantly grows
with decreasing of current. The general condition to be satisfied for maintenance of the column can be written
as $I_r\tau_e =1$, where $\tau_e$ is a mean life time of an electron ($\tau_e \cong \tau_i = \tau_e (N_i/N_e)$
at steady conditions). It is clear (and easy to see on simulation data) that lower electron density/current
discharge results in more strong radial electric field $E_r$ in bulk plasma and relatively smaller one in
sheath than in higher current case. Under the influence of this radial field the ions are accelerated towards
tube wall much faster in the lower current PC. The cause of that the mean ion velocity $v_i=\mu_i E-D_i\nabla
n_i/n_i$ is mostly determined by first term and almost does not depends on a variation of ion density profile.
Therefore the mean life time of ions/electrons is longer for higher current PC and, hence, ionization rate per
a electron must decrease with increasing of current to sustain the discharge at balance. This conclusion is
quite general but it can not directly determine the slope of voltage-current curve. Nevertheless, for the
simplest model (or for enough small current) where the ionization rate depends strongly on axial electric field
$E_z$, the equilibrium axial electric field $E_z$ have to be also less for higher current positive column
\cite{Zhakh}.

Finally we may conclude that the developed particle balance theory itself can not predict precisely the sign of
V-I slope because the axial electric field $E_z$ responds first of all for energy balance in discharge. Energy
balance equations are much more complicated and detail information about energy transfers and conversions can
be obtained by simulation only.

%
\section{Energy balance in simulated PC}

\begin{figure}
\centering
\includegraphics [width=\columnwidth] {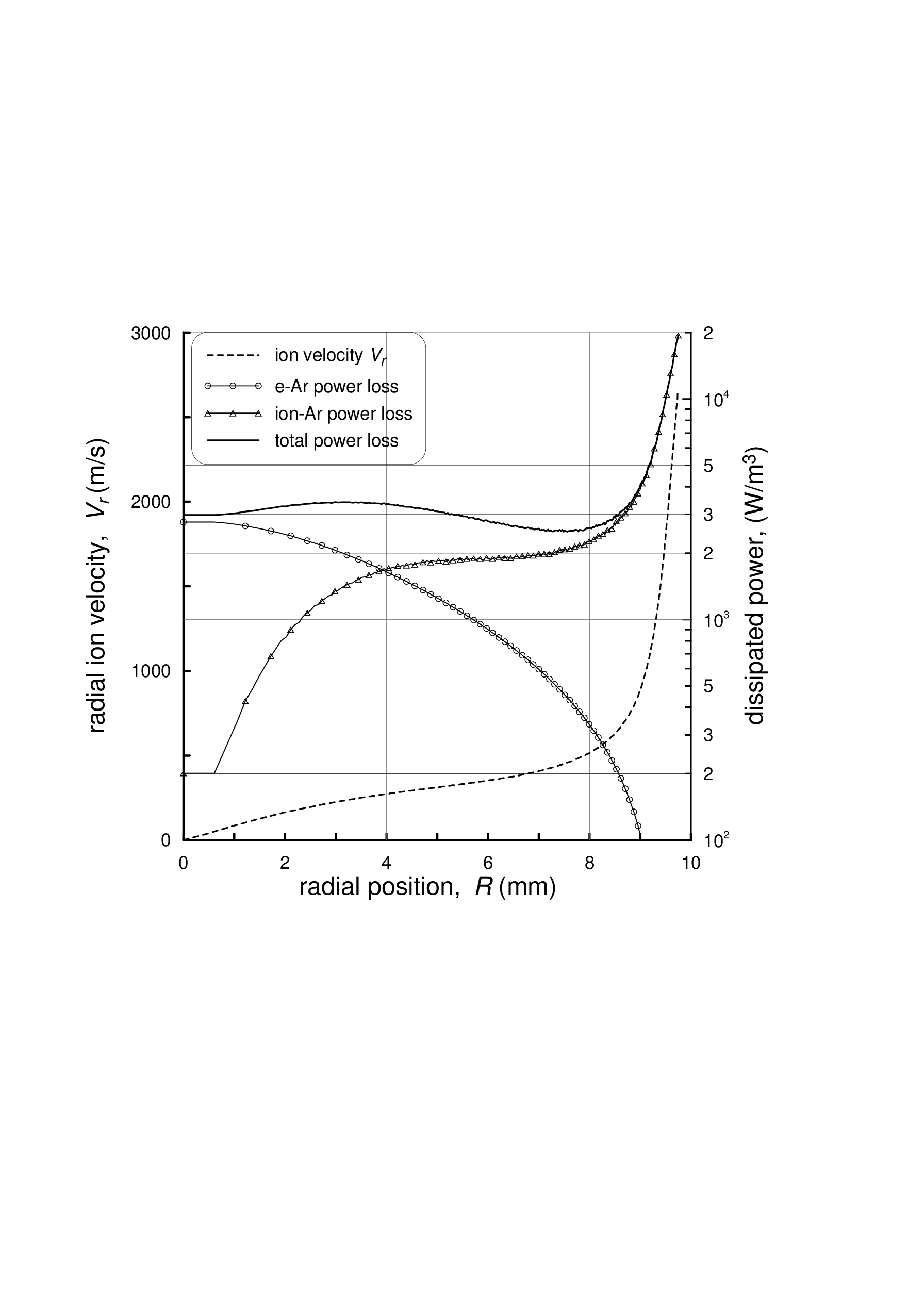}
\caption{\label{fig:ei-loss} The spatial dependency of power loss at pressure $p=0.261$ Tor
                             and current $I=50$ mA.}
\end{figure}

Electrons acquire kinetic energy from the axial electric field $E_z$, from superelastic collisions, from
excited atom-atom ionization and lose it in the electron-wall and electron-atom collisions. Ions gain kinetic
energy mostly from the radial electric field $E_r$ and lose it in the ion-atom and ion-wall collisions. Excited
atoms in their turn gain/lose energy by the electron hits and radiative transitions, and lose energy by
diffusion to the tube wall. Our MC program provides detailed information on these all processes included in the
model. According to our simulation the distribution of power loss across the tube is not uniform. Figure
\ref{fig:ei-loss} shows strong nonuniformity in the spatial dependencies of elastic electron-atom and ion-atom
power loss in pure argon discharge simulation. Elastic (and charge exchange) ion-atom collisions becomes more
important than elastic electron-atom power loss near the tube wall due to strong acceleration of ions by radial
electric field, especially in the sheath.

\begin{table}
             \caption{ Contributions of the partial power dissipations in negative V-I slope
                        $\D E_n/\D E_z$ (\%)  at gas pressure 0.261 Tor }
 \begin{center}
 \begin{tabular}{|c|c|c|c|c|}
  \hline
  Current, mA &  Diffusion & Ionization  &   Radiation    & ion-Ar col.   \\
 \hline
 $0.02 - .05$ &  13.5      &   22.5      &   30.0         &  27.8               \\
 \hline
 $0.05 - 0.1$ &  22.9      &   18.6      &   31.5         &  23.2                \\
 \hline
 $0.1  - 0.2$ &  29.2      &   16.2      &   31.9         &  20.4                \\
 \hline
 $0.2  - 0.5$ &  43.4      &   14.2      &   24.0         &  17.7               \\
 \hline
 $0.5 -  1.0$ &  54.2      &   14.5      &   13.7         &  18.4                \\
 \hline
 $1.0 -  3.0$ &  50.6      &   13.1      &   21.7         &  15.8               \\
 \hline
  $3.0 - 10$  &  31.7      &   13.3      &   40.4         &  20.5               \\
 \hline
   $10 - 30$  &   5.0      &    7.1      &   79.3         &  10.9               \\
 \hline
   $30 - 100$ &   0.8      &    3.7      &   89.3         &  7.0                \\
 \hline
 \end{tabular}
 \end{center}
\end{table}

It is evident that the input electric power must be qual to the total wasted power at a steady positive column.
Energy balance equation can be written as follow:
\begin{equation}
     W = IE_z = \sum_n W_n                            \label{eq:Eb01}
\end{equation}

The program calculates independently the input electric power $W=IE_z$ and the dissipated powers $W_n$ for the
each energetic n-channel such as light emission, electron-atom and ion-atom elastic collisions, ionization,
diffusion of excited atoms to the tube wall, electron-wall and ion-wall losses. Simulation shows that the power
balance maintains with a good accuracy. To reveal the influence of dissipating n-channel on the axial electric
field $E_z$ we may rewrite Eq.(\ref{eq:Eb01}) as
\begin{equation}
     E_z = \sum_n W_n/I = \sum_n E_n                   \label{eq:Eb02}
\end{equation}
where $E_n$ is a partial axial field which corresponds to n-channel of the total power dissipation. By using
the partial electric field $E_n$ we may easily estimate contribution of each n-channel to the total energy loss
in a steady PC and, moreover, determine the real atomic reasons behind variation of measured axial field $E_z$.

The MC program is able to predict the experimental voltage-pressure and voltage-current curve (see Figures
\ref{fig:EI_0.261Tor}-\ref{fig:EP_50mA}). The negative slope $\D E_z$ in V-I curve on Fig.\ref{fig:EI_0.261Tor}
can be decompose on partial components $\D E_n$ of total electric field $E_z$ according to equality $1 = \sum_n
\D E_n/\D E_z$ from Eq.(\ref{eq:Eb02}). As it is shown in Table I the most considerable (not for $E_z$ but for
derivative $dE_z/dI$) power loss channel is diffusion of metastable atoms to the tube wall in the case of low
current ($0.2-3$mA). For the high currents the variation of radiation power loss makes a main contribution to
the amplitude of negative slope.

The Figures \ref{fig:EI_0.261Tor}-\ref{fig:EP_50mA} also show radiative efficiency of the lamp measured as a
ratio of radiated power to the total dissipated electric power. The maximum of efficiency coincides with the
maximum of the voltage-pressure curve at pressure around 0.2 Tor and near 20 mA current.

\begin{acknowledgments}
We wish to acknowledge the support from the Lighting Research Laboratory,
              Matsushita Electric Industrial Co.
\end{acknowledgments}

\newpage 

\end{document}